# Solving mathematical problems with quantum search algorithm

**Quantum computation has attracted much attention since it was shown by Shor and Grover the possibility to implement quantum algorithms able to realize, respectively, factoring and searching in a faster way than any other known classical algorithm. It is possible to use Grover's algorithm, taking profit of its ability to find a specific value in a unordered database, to find, for example, the zero of a logical function; the minimal or maximal value in a database or to recognize if an odd number is prime or not. Here we show quantum algorithms to solve those cited mathematical problems. The solution requires the use of a quantum bit string comparator being used as oracle. This quantum circuit compares two quantum states and identifies if they are equal or, otherwise, which of them is the largest. Moreover, we also show the quantum bit string comparator allow us to implement conditional statements in quantum computation, a fundamental structure for designing of algorithms.**

The Grover's quantum search algorithm is a celebrated result in quantum computation that proves that quantum information properties (superposition) can improve the speedup of finding a specific value within an unordered database. In this case, no technique using data structures can be used and only sequential tentative can be realized. Computationally, the quantum search is proved to get approximately $O(N^{1/2})$ operations (in comparison with the $O(N)$ classical operations), which indicates a quadratic speedup[1-3]. Even though this improvement can be considered minor than other quantum algorithms, Shor[4] and Deutsch-Jozsa[5] algorithms are exponentially better than their classical alternatives, the fact is that searching is fundamental in computer science having a large amount of applications. In addition, no classical algorithm can be more efficient than Grover algorithm, that is, the quantum search algorithm is as efficient as the best search algorithm could be. The basic reason that allows this performance is the smart use of the quantum superposition which means that all states can be processed at once (in contrast with the combinatory explosion of the classical alternatives). Basically, during the processing, the database, initially an equally weighted superposition of all possible states, converges to a state that can also be a superposition, but containing only the states that are solutions of the problem, named marked states. There are several works on variations of Grover' algortihm[6,7], entanglement measures based on the Grover' algorithm[8,9] and implementation of Grover's algorithm[10,11]. Here, our goal is to show how to solve some interesting mathematical problems using the Grover's algorithm with an oracle based in a quantum circuit that compares two quantum states, representing binary strings, and identifies if they are equal or not and, in this last case, which of them is the largest (or the lowest). The circuit that makes the comparison is named quantum bit string comparator, QBSC.

**Quantum bit string comparator**

The quantum circuit shown in Fig. 1 is able to compare two binary strings (having the same number of bits) identifying, by the measurement of two qubits, if they are equal or, if they are different, which of them is the largest (or the lowest). The quantum circuit proposed makes the comparison of two strings of three bits, but the generalization to any number of bits is straightforward. Basically, the quantum circuit compare the strings bit-to-bit from the left (most significant bit) to the right (less significant bit). In a measurement of the outputs ($O_1$ and $O_2$), if $O_1=1$ and $O_2=0$ then $a>b$; if $O_1=0$ and $O_2=1$ then $a<b$; at last, if $O_1=0$ and $O_2=0$ then $a=b$. Initially, the comparison between the first bit of each string is dominant, that is, if they are different, then the outputs will be $O_1=a_1$ and $O_2=b_1$. If they are equal ($a_1=b_1$) the comparison between the second bit of each string will be dominant, that is, if they are different, then the outputs will be $O_1=a_2$ and $O_2=b_2$. If the second bits are also equal, the comparison between the third bit of each string will be dominant and so on. In the circuit of Fig. 1 the transfer of dominion from one position of the string to the next is realized by the Toffoli gate $C_3$ (activated in zero) and the CNOTS gates $C_4$ and $C_5$. Obviously, only the less significant bit does not have the dominion transfer circuit. The following examples will make clear the functioning of the circuit (for simplicity it will be considered the comparison of two states of two bits, but the result is directly generalized for any number of bits).

In Table I, the number inside the parenthesis besides the qubit means the probability of the output to be that qubit. For example, comparing $|a\rangle=|11\rangle$ with $|b\rangle=\alpha|01\rangle+\beta|11\rangle$, with probability $|\alpha|^2$ $|a\rangle >|b\rangle$ and, hence, $|O_1O_2\rangle=|10\rangle$. On the other hand, with probability $|\beta|^2$ $|a\rangle=|b\rangle$ and, hence, $|O_1O_2\rangle=|00\rangle$.

## Searching for a minimal value in a database

There are several important applications of the comparison of binary strings in quantum computation. Let us firstly discuss the use of the QBSC as an oracle in the Grover's quantum search algorithm. An example of 4 qubits is shown in Fig. 2, but a generalization for any number of qubits is straightforward.

For the circuit shown in Fig. 2, the task is to search in the database words of four bits larger than "0111". This is clearly a typical case of multiple marked states and, hence, the output of Grover's algorithm will be a superposition of the all possible solutions. The reference state $|0111\rangle$ works as string $|a\rangle$ while the state of Grover's algorithm is $|b\rangle$. If the initial state of Grover's algorithm is $(1/4)\sum_{i=0}^{15}|i\rangle$ then the output state will be:

$$\frac{(|8\rangle+|9\rangle+|10\rangle+|11\rangle+|12\rangle+|13\rangle+|14\rangle+|15\rangle)}{2\sqrt{2}} \qquad (1)$$

where the decimal representation has been used for simplification. If instead of search for states larger than $|0111\rangle$ one was looking for states lower than $|0111\rangle$, then $O_2$ would be used instead of $O_1$ (the reference is $|a\rangle$). Let us now suppose that the goal is to find the minimal value in the database. In order to find the minimal value the quantum circuit shown in Fig. 3 has to be used to activate the lowest CNOT of Grover's quantum circuit:

Using the circuit of Fig. 3, the oracle will recognize strings lower or equal than the reference. The algorithm to find the minimum is as follows: Initially, one value of the database is randomly chosen. This value will be used for comparison (string $|a\rangle$). The algorithm runs and, at end, the result of the measurement will be one of the members of the database lower or equal than the initial value used. The result of the measurement will now be used as the new value to be compared. The process is repeated till the result of the measurement does not change anymore.

## Finding the zero of a logical function

Another application is the interesting problem of finding the zero of a function. Let us assume that the unitary transformation $U_f$ represents the function whose zero one wish to find. The following quantum circuit as an oracle in the Grove algorithm will solve the problem.

In Fig. 4, $x$ is a binary number of $n$ bits. The quantum comparator will activate the CNOT ($O_1=O_2=0$) only if $|f(x)\rangle=|0^{\otimes n}\rangle$, hence, the oracle will recognize only the zero of the function $f$. Changing the reference state (and not the oracle circuit), it is possible to find any desired value.

## Detecting a prime number

Now, let us move to a third problem. We wish to decide if $a$, an odd number, is a prime number or not. In order to answer this question, the following quantum circuit can be used:

The quantum circuit of Fig. 5 works as follows: the number to be tested is coded in the quantum state $|a\rangle$ whereas $|b\rangle$ is, as usual, the equally weighted superposition of all possible states having $n$ bits. The first Grover/QBSC algorithm has as output $|b_1\rangle$, the equally weighted superposition of all possible states having $n$ bits lower than $a$. This state passes for the second Grover algorithm whose oracle will recognize as marked states only the odd numbers. This is the database $|b_{o1}\rangle$ that will be used. Since two copies of database are necessaries, there exist two Database preparation blocks. Both database are inputs of a quantum circuit, block PROD, which gives as output the product of its inputs. For example, having inputs $|a\rangle$, $|b\rangle$ and ancilla $|0\rangle$, the output of PROD is $|a\rangle|b\rangle|a\mathrm{x}b\rangle$. A quantum circuit for multiplication can be designed as its classical counterpart. A third Grover's algorithm is used for searching in the database $|b_{o1}b_{o2}\rangle$ a state equal to $|a\rangle$. For this task, the oracle QBSC is once more used. If there exist one, then $a$ is not a prime number. In order to check this, at the end of the last quantum search algorithm, the state $|b_f\rangle$ is measured and the value obtained is compared to *a*. If both are equal, then *a* is not prime, otherwise, *a* is a prime number. As an example let us assume *a*=17 (decimal). In this case one has (unnormalized states):

$$|b_1\rangle = \sum_{i=0}^{16}|i\rangle \rightarrow |b_{o1(2)}\rangle = |1\rangle + |3\rangle + |5\rangle + |7\rangle + |9\rangle + |11\rangle + |13\rangle + |15\rangle \rightarrow \quad (2)$$

$$\rightarrow |b_{o1}b_{o2}\rangle = |b_o\rangle + |3b_o\rangle + |5b_o\rangle + |7b_o\rangle + |9b_o\rangle + |11b_o\rangle + |13b_o\rangle + |15b_o\rangle \quad (3)$$

$$|nb_o\rangle = |n\rangle + |3n\rangle + |5n\rangle + |7n\rangle + |9n\rangle + |11n\rangle + |13n\rangle + |15n\rangle, \ n=1,3,...,15 \quad (4)$$

Since 17 is different of |$nb_o$⟩ for any *n*, in a measurement of |$b_f$⟩ a random value will be measured and, since 17 is prime, the value measured will be different of it. On the other hand, if *a* was for example 15, the Grover algorithm with oracle QBSC, would find it in the database |$b_{o1}b_{o2}$⟩, and the result $b_f$ measured would be equal to *a*. Still considering Fig. 7, one easily sees that the complete output of the last Grover algorithm is, for a non-prime number *a*, |$b_{o1}$⟩|$b_{o2}$⟩|$a$⟩, where $a=b_{o1}b_{o2}$. Now one can test if $b_{o1}$ and $b_{o2}$ are themselves primes or not. Repeating this process, one can obtain all prime factors of *a*.

## Conditional statements in quantum computation

Now we can invert the situation using the QBSC to control the oracle of a Grover algorithm. In this case, the application of the quantum comparator circuit is the realization of conditional statements as, for example, If *a*>*b* then do "something". For instance, using again the Grover algorithm as scenario, we can implement the following piece of software: If |*a*⟩>|*b*⟩ then search for solution $S_1$, otherwise, search for solution $S_2$. For an oracle based on *n*-CNOTS, the quantum circuit for a four bit problem is:

In Fig. 6, if *a*>*b* the Grover algorithm will search for |0011⟩, otherwise the algorithm will search for |1100⟩. Other choices are possible using combinations of both outputs $O_1$ and $O_2$.

# Discussions

The quantum bit string comparator enables the implementation of quantum algorithm using conditional statements, a fundamental structure for designing of algorithms. This enlarges the number of applications where quantum algorithms can be used and, at the same time, it brings close to quantum programmers successful techniques used in classical computation based on comparisons. Furthermore, the use of the QBSC with Grover algorithm gives us power to solve some mathematical problems of the type presented in this work, as well open the possibility to create quantum algorithm with very specific tasks. For example, constructing a database composed only of prime numbers it is possible, using the QBSC and Grover algorithm, to search for an even number that does not satisfy Goldbach's conjecture (all even number larger than two can be written as the sum of two prime numbers).

**Acknowledgements**

This work was supported by the Brazilian agency FUNCAP.

**Authors Information**

Rubens Viana Ramos, Paulo Benício Melo de Sousa and David Sena Oliveira

Department of Teleinformatic Engineering, Federal University of Ceara, DETI/UFC C.P. 6007, Campus do Pici, 60755-640 Fortaleza-Ce Brazil.

**Correspondence** and requests for materials should be addressed to rubens@deti.ufc.br.

TABLE I

| $|a\rangle$ | $|10\rangle$ | $|00\rangle$ | $|10\rangle$ | $|11\rangle$ | $|01\rangle$ |
|---|---|---|---|---|---|
| $|b\rangle$ | $\alpha|00\rangle+\beta|01\rangle$ | $\alpha|01\rangle+\beta|10\rangle$ | $\alpha|00\rangle+\beta|11\rangle$ | $\alpha|01\rangle+\beta|11\rangle$ | $\alpha|01\rangle+\beta|11\rangle$ |
| $O_1$ | $|1\rangle$ (1) | $|0\rangle$ (1) | $|1\rangle$ ($|\alpha|^2$) | $|1\rangle$ ($|\alpha|^2$) | $|0\rangle$ (1) |
| $O_2$ | $|0\rangle$ (1) | $|1\rangle$ (1) | $|1\rangle$ ($|\beta|^2$) | $|0\rangle$ (1) | $|1\rangle$ ($|\beta|^2$) |

FIGURE 1

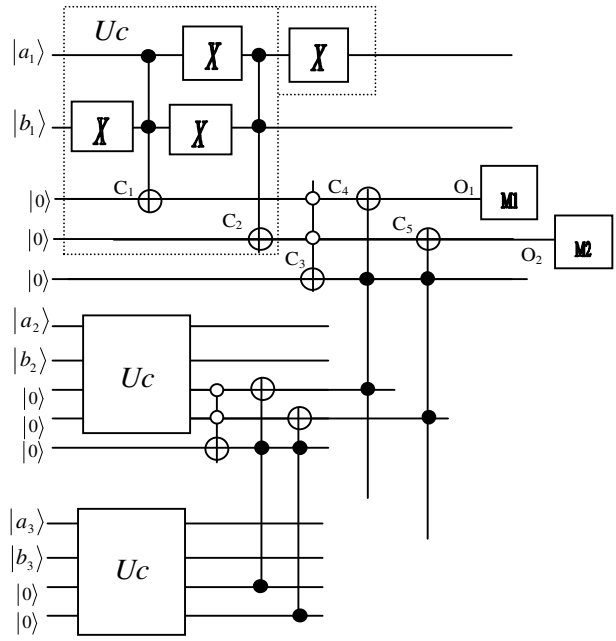

FIGURE 2

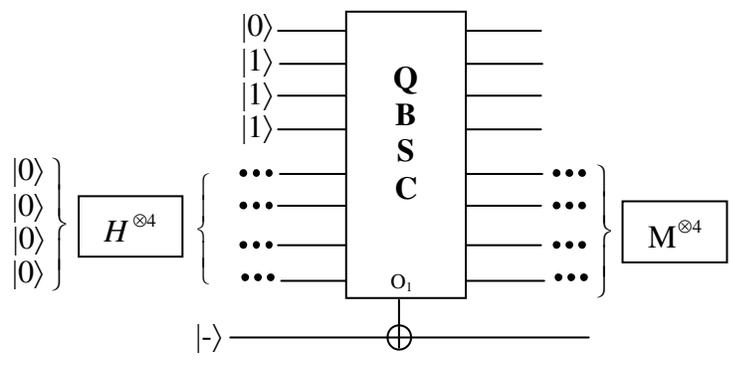

FIGURE 3

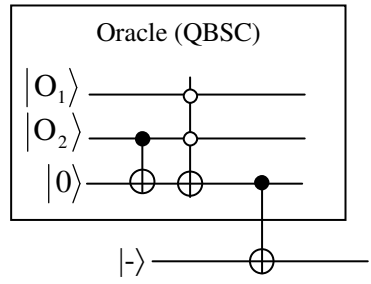

FIGURE 4

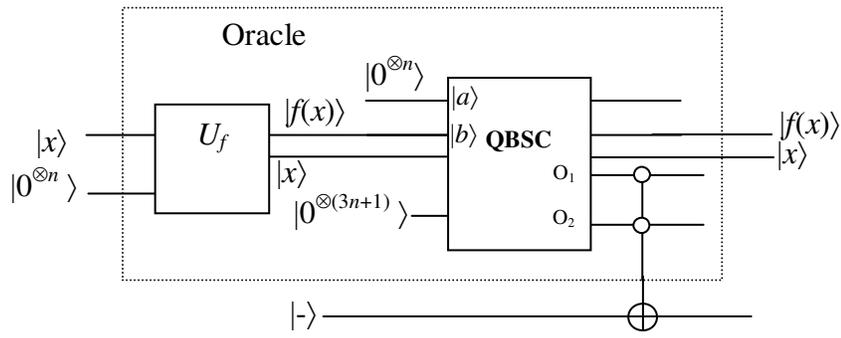

FIGURE 5

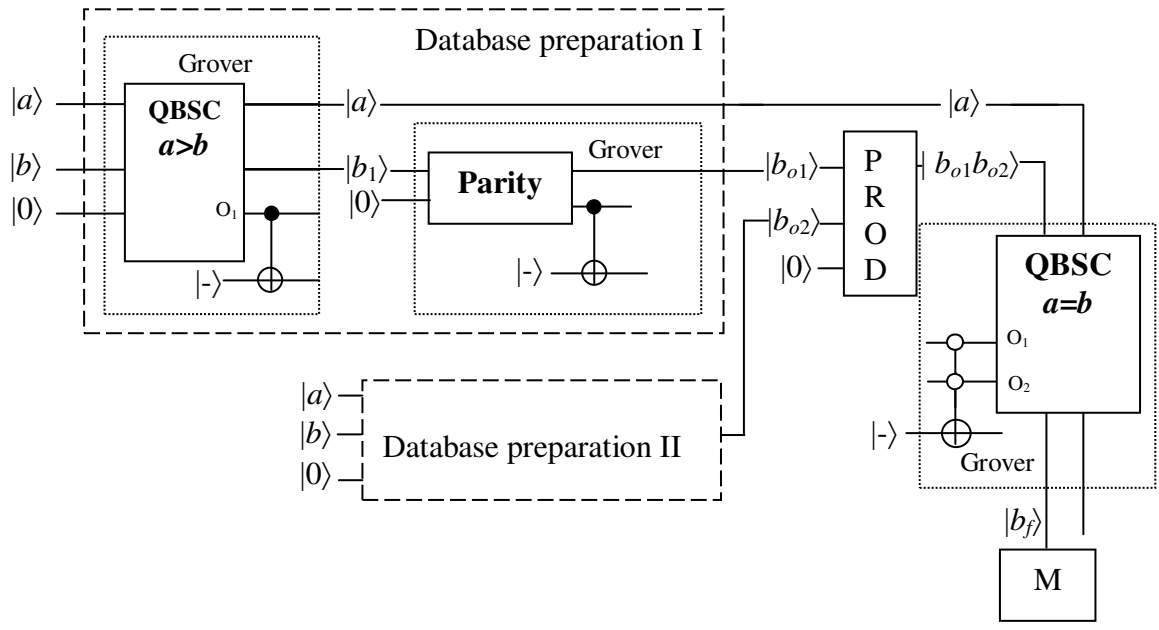

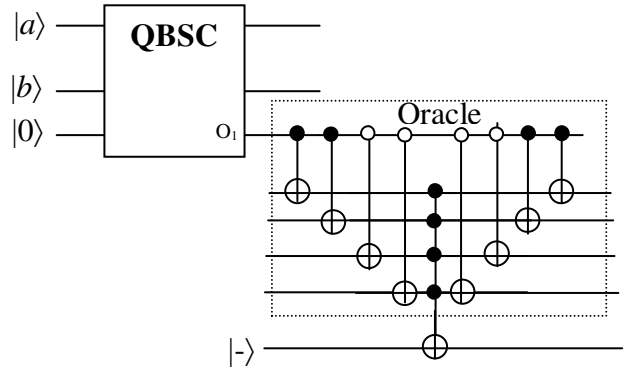

Legend of Tables

Table I. Examples of the output of the quantum bit string comparator for two strings of two qubits at the inputs.

Legend of Figures

**Figure 1 | Quantum circuit for comparison of two strings of three qubits: $|a\rangle=|a_1\rangle|a_2\rangle|a_3\rangle$ and $|b\rangle=|b_1\rangle|b_2\rangle|b_3\rangle$. $M_1$ and $M_2$ are measurers.** The quantum circuit $U_c$ is reponsible for determination if both compared qubits are equal or not. If they are differente their values will apear at the outputs $O_1$ and $O_2$, otherwise, the quantum circuit composed by a Tofolli gate (activated in zero) and two CNOTs are responsible for transfer the decision to the following par of bits of the strings.

**Figure 2 | QBSC circuit as an oracle in a Grover search algorithm of four qubits. $H$- Hadamard gate. M – Measurer.** Quantum bit string comparator working as an oracle in a Grover algorithm will recognize as marked states those states that are larger than $|0111\rangle$, information provided by output $O_1$. The QBSC ancillas are not shown.

**Figure 3 | Control of the lowest CNOT of Grover's quantum circuit in order to find the minimal value in a database using the quantum comparator circuit as oracle.** In order to find the minimal value in the database, the oracle will recognize as marked states those states that are lower or equal to the reference state, $|a\rangle$.

**Figure 4 | Quantum circuit for finding the zero of a logical function represented by $U_f$, using Grover algorithm.** The quantum bit string comparator, working as oracle, will recognize when the output of the function is equal to the reference state, $|0^{\otimes n}\rangle$. The ancillas used by the QBSC form the state $|0^{\otimes(3n+1)}\rangle$.

**Figure 5 | Quantum circuit for determination if an odd number is prime or not.** In order to test if $a$ is prime or not, it is necessary the tensor product of two states where each one of them ($|b_o\rangle$) is the superposition of all odd number states lower than $a$. Using a multiplier (PROD) having the tensor product $|b_o\rangle|b_o\rangle$ as input, it is possible to produce all possible products of two odd numbers lower than $a$. At last, it is necessary only to compare $a$ with those products, through Grover's algorithm having a QBSC as oracle, in order to check if $a$ is divisible for any odd number lower than its value.

**Figure 6 | Implementation of the conditional statement: If $a>b$ then search for solution $S_1=|0011\rangle$, otherwise, search for solution $S_2=|1100\rangle$.** The result of the comparison between $|a\rangle$ and $|b\rangle$ control the set of CNOTgates that will be enabled to invert their target qubit.